\documentclass[10pt,aps,prd,nofootinbib,superscriptaddress,twocolumn]{revtex4}
\usepackage[utf8]{inputenc}
\DeclareUnicodeCharacter{200B}{{\hskip 0pt}}
\usepackage{color}
\usepackage{graphicx}
\usepackage{amsmath}
\usepackage{amssymb}
\usepackage{bm}
\usepackage{acronym}
\usepackage{ifthen}
\usepackage{blindtext}
\usepackage[normalem]{ulem}
\usepackage{hyperref}
\usepackage{etoolbox}
\usepackage{fancyhdr}
\usepackage{xspace}
\usepackage{textcomp}
\usepackage{multirow}
\usepackage[caption=false]{subfig}
\usepackage{lineno}
\usepackage{tabularx}
\hypersetup{
    colorlinks=true,
    linkcolor=blue,
    filecolor=magenta,      
    urlcolor=black,
		citecolor=red,
		}

\begin{document}

\title{Blueshift of light rays induced by gravitational wave memory effect}

\author{F. L. Carneiro}
\email{fernandolessa45@gmail.com}
\affiliation{Departamento de F\'isica, Universidade Federal do Norte do Tocantins, 77824-838, Aragua\'ina, TO, Brazil}

\author{S. C. Ulhoa}
\email{sc.ulhoa@gmail.com}
\affiliation{Instituto de F\'isica, Universidade de Bras\'ilia, 70910-900, Bras\'ilia, DF, Brazil}
\affiliation{Canadian Quantum Research Center,\\ 
204-3002 32 Ave Vernon, BC V1T 2L7  Canada}

\author{J. W. Maluf}
\email{jwmaluf@gmail.com}
\affiliation{Instituto de F\'isica, Universidade de Bras\'ilia, 70910-900, Bras\'ilia, DF, Brazil}

\date{\today}

\begin{abstract}
The article deals with photon propagation in pp-wave spacetimes in the strong gravitational-wave regime and its consequences for redshift measurements. We show that null geodesics crossing a localized pp-wave pulse exhibit an energy memory effect, producing a finite asymptotic shift in the photon frequency measured by static observers. This path-dependent contribution acts directly on the redshift observable and may help account for divergent interpretations of supernova redshift data in the presence of intervening gravitational radiation.
\end{abstract}

\maketitle

\section{Introduction}
The accelerated expansion of the universe, first inferred in the late
1990s from type Ia supernova observations \cite{Riess1998,Perlmutter1999},
has in recent years been the subject of increasingly explicit criticism
\cite{Nielsen2016,Tutusaus2017,Tutusaus2019,Colin2019,Mohayaee2021}.
Although subsequent analyses based on larger and more refined data sets
have largely confirmed the presence of accelerated expansion
\cite{Betoule2014,Scolnic2018}, a persistent tension has emerged within
the community.

This apparent divide does not arise solely from the use of different
observational techniques, but mainly from the accumulation and
cross-comparison of increasingly large data sets. A common feature shared
by all these methods is their reliance on the apparent change in the
frequency of light emitted by distant sources, interpreted as a
consequence of their relative motion. In this context, standard candles
are employed to compare relative motions and to infer the presence of
cosmic acceleration.

Implicit in this procedure is the assumption that no physical mechanism
other than cosmological expansion is capable of producing a cumulative
modification of the observed photon frequency, apart from well-known and
controllable sources of noise. It is therefore natural to ask whether the
divergences observed in the determination of cosmic acceleration may be
related to additional physical mechanisms capable of altering the
frequency of light emitted by standard candles during its propagation.
As more observational methods and data sets are combined, the probability
of including measurements affected by such mechanisms necessarily
increases.

In this work, we investigate one possible mechanism that has not been
systematically accounted for so far.
Throughout this work, localized gravitational radiation is modeled by
pp-wave pulses, which provide an exact and analytically tractable
description of gravitational waves propagating over finite intervals.
While idealized, pp-waves capture the essential nonlinear features
relevant to memory effects and allow for a clean and unambiguous analysis
of photon propagation through gravitational radiation.
In the following, we show how the
interaction of a null geodesic with a pp-wave spacetime gives rise to a
memory effect, leading to a finite asymptotic shift in the photon
frequency.

Plain-fronted gravitational waves with parallel rays (pp-waves) are defined by an
expansion-, shear-, and rotation-free congruence of null geodesics that admits
a null Killing vector $k_{\mu}$ along its propagation direction. By constructing
a double-null coordinate system, in which $k_{\mu}$ is tangent to the null
coordinate $v$ and an auxiliary null vector $N_{\mu}$ is tangent to the other
null coordinate $u$, the pp-wave line element may be written in Brinkmann
coordinates $x^{\mu}=\{u,x,y,v\}$ as
\begin{equation}\label{eq1}
	ds^{2}=H(u,x,y)\,du^{2}+dx^{2}+dy^{2}-2\,du\,dv\,,
\end{equation}
where $x$ and $y$ denote the two-dimensional transverse space orthogonal to the
null propagation direction $v$, and the profile $H(u,x,y)=f(u)F(x,y)$ satisfies
the Laplace equation
\begin{equation}
(\partial_x^2+\partial_y^2)H(u,x,y)=0\,.
\end{equation}
The function $f(u)$ characterizes the pulse profile of the wave. The solutions
of the two-dimensional Laplace equation can be written as
\begin{equation}
H(u,x+iy)=f(u)(x+iy)^m\,,
\end{equation}
with $m=2,3,4,\ldots$, giving rise to the polarization modes
\begin{align}
	H(u,x,y) &= f(u)\,\mathrm{Re}\left[(x+iy)^m\right]\,,\nonumber\\
	H(u,x,y) &= f(u)\,\mathrm{Im}\left[(x+iy)^m\right]\,,\label{polarizations}
\end{align}
of the pp-waves. The lowest-order modes, corresponding to $m=2$, are given by
\begin{align}
	H_{+}(u,x,y) &= f(u)\,(x^{2}-y^{2})\,,\label{plus}\\
	H_{\times}(u,x,y) &= f(u)\,(x\,y)\,,\label{times}
\end{align}
which correspond to the polarization modes of linearized gravitational waves.

The modes given by Eqs.~\eqref{plus} and \eqref{times} yield curvature tensor
components that are uniform in the transverse plane, thereby producing a
uniform geodesic deviation along the wavefront. In contrast, higher-order modes
exhibit curvature components that depend explicitly on the transverse
coordinates $(x,y)$, generating non-uniform patterns of geodesic deviation for
test particles.

When interacting with test particles, the nonlinearity of pp-waves causes a
permanent change in the relative separation of initially neighboring worldlines
between two asymptotic regions. This phenomenon is known as the memory effect
\cite{zhang2017memory,zhang2017soft,zhang2018memory,zhao2025gravitational}. In addition to altering
relative distances, the wave can also induce a permanent change in the relative
velocities of particles, giving rise to a velocity memory effect
\cite{zhang2018velocity,zhang2024displacement}. As a consequence of this permanent modification of the
kinematical state, the particle energy after the interaction may differ from
its energy before the interaction, implying an exchange of energy between the
wave and the particles. Depending on the initial conditions and on the wave
profile, the particle energy may either increase or decrease
\cite{maluf2018plane,maluf2018variations,abbasi2025kinetic}.

The interaction of pp-waves with light rays propagating parallel to the positive
$v$ direction produces no memory effect, since the expansion-, shear-, and
rotation-free nature of the null congruence ensures that initially parallel
light rays remain parallel during and after the interaction.

However, if the light rays are not parallel to the propagation direction of the
pp-wave, e.g., if they possess components in the transverse plane or
propagate parallel to the wave but in the opposite direction, initially parallel
rays will generally fail to remain parallel after a finite interaction time,
thereby retaining a memory of the interaction with the wave.

In this letter, we investigate the memory effect induced by pp-waves on light
rays approaching from arbitrary directions. By numerically solving the null
geodesic equations, we evaluate the influence of the wave on the propagation of
light rays and show that, except for the special case in which the rays
propagate parallel and in the same direction as the wave, a memory effect is
always present, as discussed in Sec.~\ref{sec2}.

In order to assess whether pp-waves can also alter photon energy, we analyze the
evolution of the photon energy as a function of the null coordinate $u$.
Adopting a coordinate-invariant approach, we construct a set of tetrads
associated with the pp-wave spacetime and adapted to a stationary observer, and
project the photon four-momentum $p^{\mu}$ onto the observer frame in
Sec.~\ref{sec3}. By considering several initial conditions, we show that the
photon energy may either increase or decrease depending on its trajectory, with
an overall tendency toward energy gain when multiple interactions are taken into
account. Finally, we discuss the potential implications of this cumulative
contribution for light rays propagating over cosmological distances, such as
those originating from distant galaxies. Since precise redshift measurements are
fundamental to modern cosmology, our results indicate that interactions between
light rays and gravitational waves may, through the memory effect, introduce
path-dependent quantitative corrections to the interpretation of observational
redshift data.

\section{Memory effect of light rays in pp-waves}\label{sec2}

For pp-waves written in Brinkmann coordinates, the transverse null geodesic
equations are
\begin{equation}\label{geotrans}
\ddot{x} = \frac{1}{2}\,\partial_{x}H\,,
\qquad
\ddot{y} = \frac{1}{2}\,\partial_{y}H\,,
\end{equation}
while the null condition $g_{\mu\nu}\dot{x}^{\mu}\dot{x}^{\nu}=0$ yields
\begin{equation}\label{geolong}
	\dot{v}=\frac{1}{2}\bigl(H + \dot{x}^{2} + \dot{y}^{2}\bigr)\,.
\end{equation}
Dots denote differentiation with respect to the null coordinate $u$, chosen
as an affine parameter ($\dot{u}=1$). This parametrization is regular for all
null geodesics except those propagating exactly parallel to the positive
propagation direction of the wave, which do not exhibit any memory effect.

The null condition prevents the independent specification of all three
velocity components. We therefore fix the initial transverse velocities
$\dot{x}_{0}$ and $\dot{y}_{0}$, where the subscript $0$ denotes quantities
evaluated before the interaction with the wave, i.e., in the asymptotic
region $u_{0}\rightarrow -\infty$ (numerically, $u_{0}=-10$).

The gravitational wave pulse is modeled by a Gaussian profile,
\begin{equation}
	f(u)=A\,e^{-u^{2}/\Lambda^{2}}\,,
\end{equation}
with amplitude $A$ and width $\Lambda$. Unless stated otherwise, we fix
$A=10^{-1}$ and $\Lambda=1$.
The amplitude $A$ of the pp-wave is a free parameter chosen to illustrate the
energy memory effect on photons; it does not correspond to any measured
astrophysical strain. Larger amplitudes are used to explore situations in
which light traverses regions of relatively strong gravitational waves,
making the effect manifest.

Evaluating Eq.~\eqref{geolong} in the asymptotic region where $H=0$, one finds
\begin{equation}
	\dot{v}_{0}=\frac{1}{2}\bigl(\dot{x}_{0}^{2} + \dot{y}_{0}^{2}\bigr)\,.
\end{equation}
The null coordinates $(u,v)$ are related to the Cartesian-like coordinates
$(t,z)$ by
\begin{equation}\label{cartesians}
u = \frac{t-z}{\sqrt{2}}\,, \qquad
v = \frac{t+z}{\sqrt{2}}\,,
\end{equation}
which implies
\begin{equation}
	\dot{z}_{0}
	=\frac{1}{2\sqrt{2}}\bigl(\dot{x}_{0}^{2} + \dot{y}_{0}^{2}\bigr)
	-\frac{1}{\sqrt{2}}\,.
\end{equation}
For $\dot{x}_{0}=0=\dot{y}_{0}$, one has $\dot{z}_{0}<0$, corresponding to
light rays propagating opposite to the wave direction. In contrast,
$\dot{x}_{0}^{2}+\dot{y}_{0}^{2}>2$ implies $\dot{z}_{0}>0$, while the case
$\dot{x}_{0}^{2}+\dot{y}_{0}^{2}=2$ corresponds to purely transverse
propagation.

\begin{figure}[t]
\centering
\begin{minipage}{0.48\linewidth}
\centering
\includegraphics[width=\linewidth]{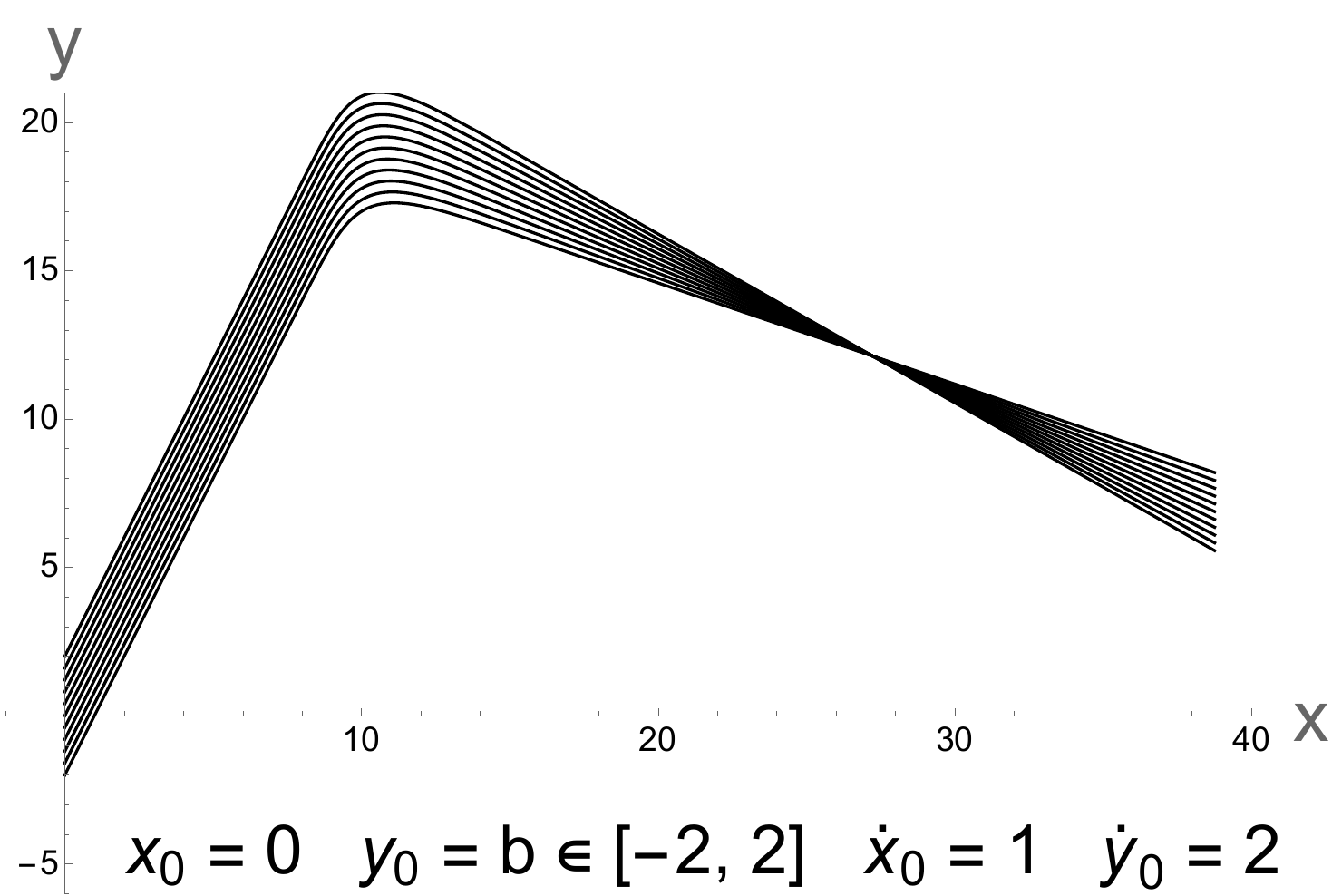}
\par\smallskip
{\small (a) Transverse view.}
\end{minipage}\hfill
\begin{minipage}{0.48\linewidth}
\centering
\includegraphics[width=\linewidth]{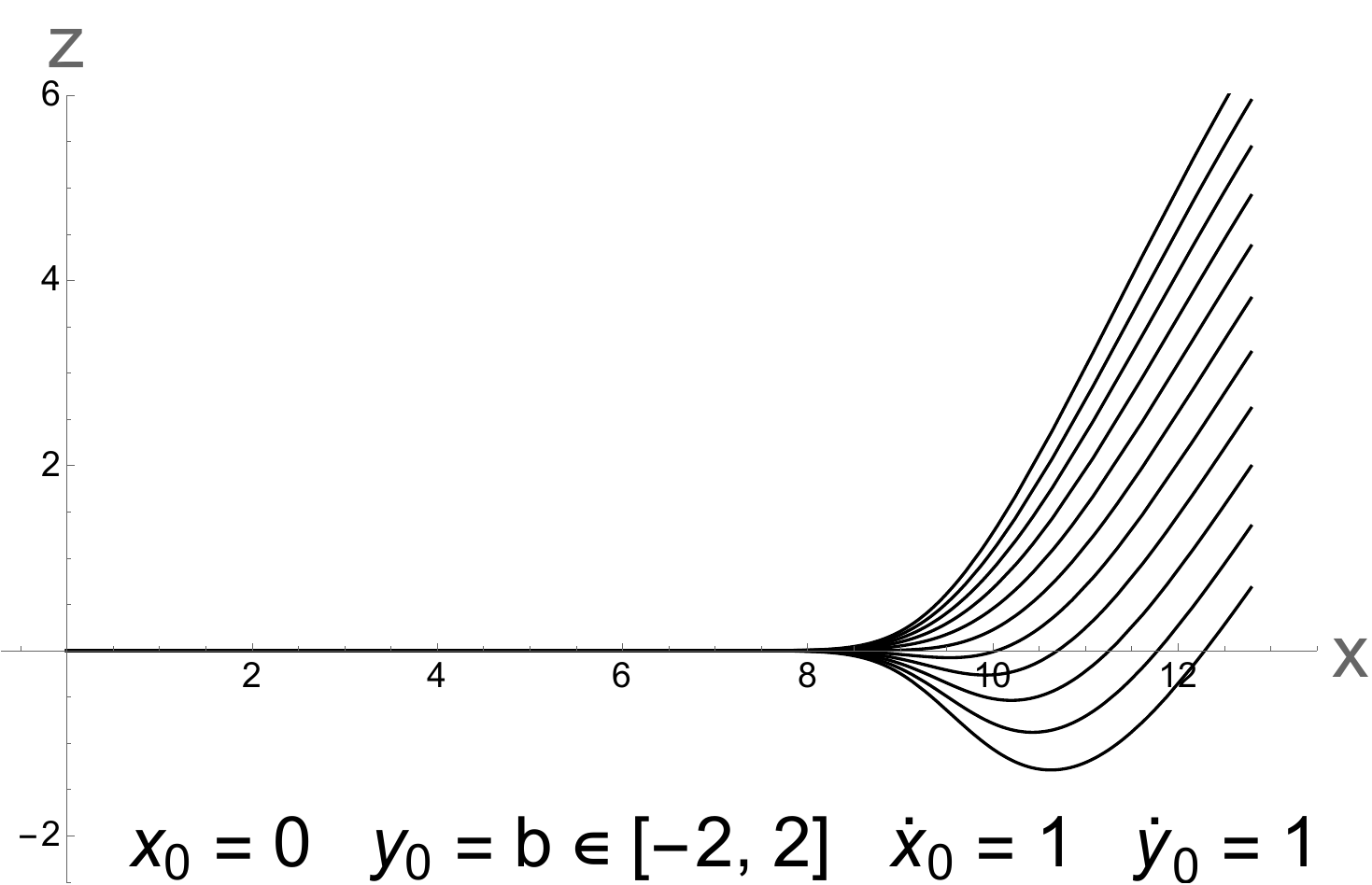}
\par\smallskip
{\small (b) Longitudinal view.}
\end{minipage}
\caption{Interaction of a beam of parallel light rays with a pp-wave of $+$
polarization.}\label{beam2D}
\end{figure}

We now illustrate the memory effect for representative configurations.
First, we consider a beam of parallel light rays with fixed initial
transverse velocities and fixed $x_{0}$, while the initial transverse
positions vary as $-b\leq y_{0}<b$. As shown in Fig.~\ref{beam2D}, the wave
induces the formation of a caustic and generates longitudinal velocity
components in an initially transverse beam.

\begin{figure}[t]
\centering
\begin{minipage}{0.48\linewidth}
\centering
\includegraphics[width=\linewidth]{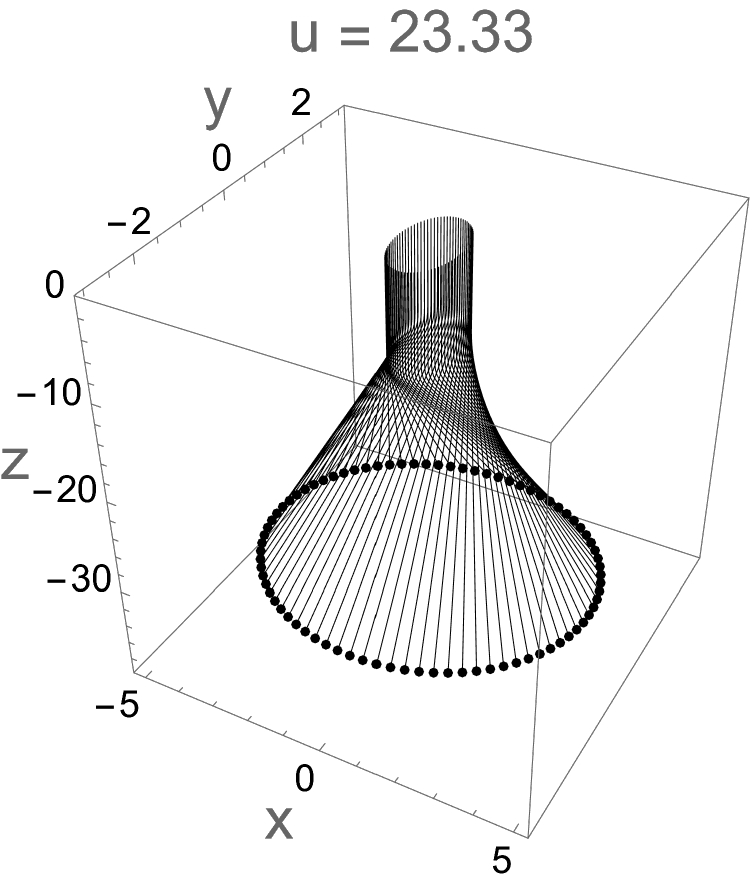}
\par\smallskip
{\small (a) $H_{+}=f(u)\,(x^{2}-y^{2})$.}
\end{minipage}\hfill
\begin{minipage}{0.48\linewidth}
\centering
\includegraphics[width=\linewidth]{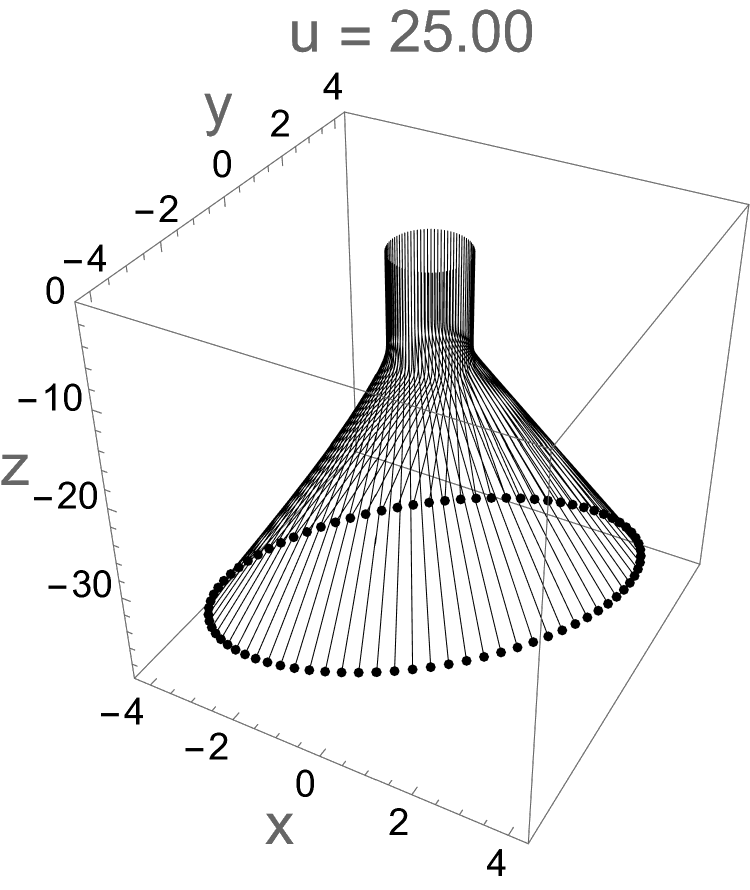}
\par\smallskip
{\small (b) $H_{\times}=f(u)\,(x\,y)$.}
\end{minipage}
\caption{Evolution of a circular beam of light rays initially propagating
opposite to the wave direction for the standard polarization modes.}
\label{beam3D}
\end{figure}

We next consider a circular beam of light rays propagating parallel to the
negative $z$ direction. In this case, the wave induces transverse motion,
deforming the initial circle into a caustic line and subsequently producing
a circular configuration with varying radius, as shown in
Fig.~\ref{beam3D}.

\begin{figure}[t]
\centering
\begin{minipage}{0.48\linewidth}
\centering
\includegraphics[width=\linewidth]{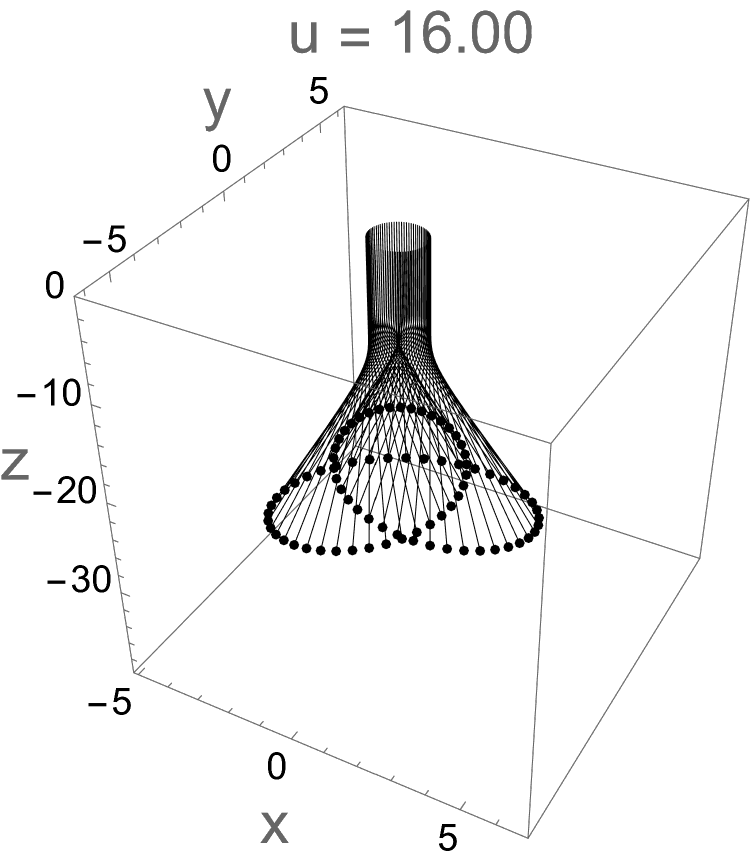}
\par\smallskip
{\small (a) $H_{3R}=f(u)(x^{3}-3\,x\,y^{2})$.}
\end{minipage}\hfill
\begin{minipage}{0.48\linewidth}
\centering
\includegraphics[width=\linewidth]{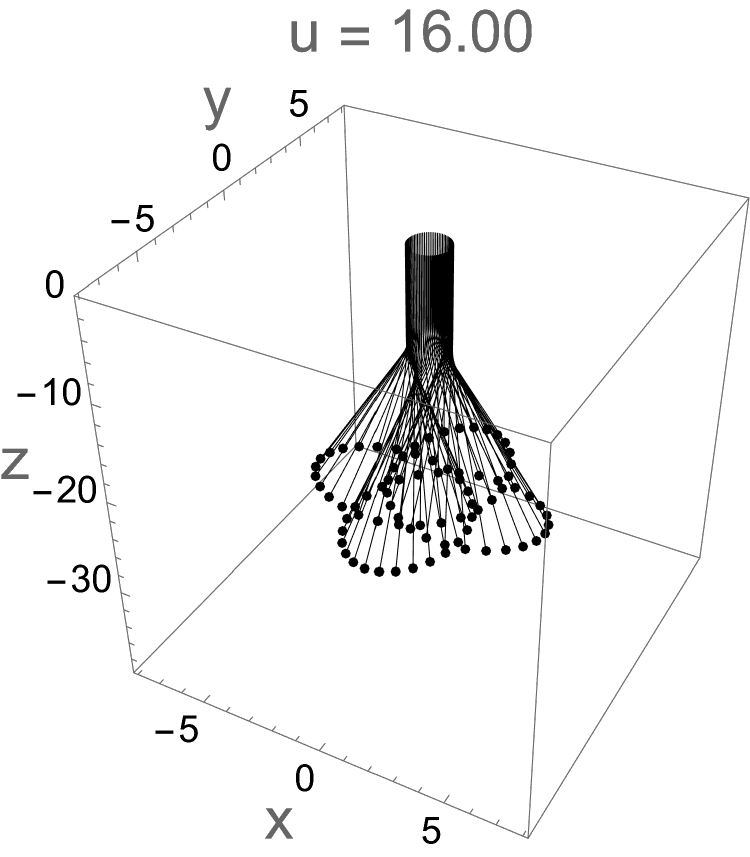}
\par\smallskip
{\small (b) $H_{4R}=f(u)(x^{4}-6\,x^{2}y^{2}+y^{4})$.}
\end{minipage}
\caption{Evolution of a circular beam of light rays initially propagating
opposite to the wave direction for higher-order polarization modes.}
\label{beam3Daliii}
\end{figure}

The evolution of circular beams also provides a geometric interpretation
of higher-order polarization modes. For instance, for $m=3$ and $m=4$ in
Eq.~\eqref{polarizations}, an initially circular configuration decomposes
into $m$ distinct structures, as illustrated in Fig.~\ref{beam3Daliii}.
This qualitative pattern persists for higher values of $m$.

Animations corresponding to these evolutions are provided as supplementary
material. In all simulations, we consider beams composed of $80$ light rays
uniformly distributed along a circle of radius $R=0.8$, with initial
transverse velocities $\dot{x}_{0}=0=\dot{y}_{0}$.

\section{Observer-measured photon energy in pp-waves}\label{sec3}

We model a light ray as being composed of test photons whose four-momentum
is defined as \cite{hobson2006general}
\begin{equation}
	p^{\mu}=\alpha\,k^{\mu}\,,
\end{equation}
where $\alpha$ is a constant and $k^{\mu}=dx^{\mu}/du$ is the tangent vector
to the photon worldline. By choosing $\alpha=1$, one obtains
\begin{equation}
	p^{\mu}=\dot{x}^{\mu}\,.
\end{equation}

In order to compute the photon energy in a coordinate-invariant manner, we
project its four-momentum onto an orthonormal frame defined by a set of
tetrads $e_{a}{}^{\mu}$. Latin indices from the beginning of the alphabet,
$a,b,\ldots=(0),(i)$, denote local $SO(3,1)$ indices associated with the
flat tangent space $T_{P}\mathcal{M}$ at an event $P$, described by the
coordinates $x^{\mu}$ of the manifold $\mathcal{M}$. Greek indices
$\mu,\nu,\ldots=0,i$ label spacetime coordinates on $\mathcal{M}$. Local
indices are raised and lowered with the Minkowski metric
$\eta_{ab}=\mathrm{diag}(-1,1,1,1)$, whereas spacetime indices are raised
and lowered with the metric tensor $g_{\mu\nu}$.

The set of four orthonormal vectors
$\{e_{(0)}{}^{\mu},e_{(1)}{}^{\mu},e_{(2)}{}^{\mu},e_{(3)}{}^{\mu}\}$
defines the instantaneous rest frame (IRF) of an observer moving along a
worldline $x^{\mu}(\tau)$, where $\tau$ denotes the observer proper time.
Since the observer is at rest in its IRF, the timelike tetrad vector must
be tangent to the worldline. Therefore, one identifies
\begin{equation}
	e_{(0)}{}^{\mu}=u^{\mu}\,,
\end{equation}
where $u^{\mu}$ is the observer four-velocity.

Physical quantities are measured by projecting them onto the IRF. In
particular, the photon energy measured by the observer is given by
\begin{equation}\label{enerproj}
	p^{(0)}=e^{(0)}{}_{\nu}p^{\nu}
	=-g_{\mu\nu}u^{\mu}k^{\nu}\,.
\end{equation}

We now construct a stationary observer associated with the pp-wave line
element~\eqref{eq1}. In Cartesian-like coordinates
$x'^{\mu}=(t,x,y,z)$, a static observer satisfies
$u'^{\mu}=(u'^{0},0,0,0)$. Transforming to double-null coordinates
$x^{\mu}=(u,x,y,v)$, the observer four-velocity becomes
\begin{equation}\label{static}
	u^{\mu}
	=\frac{\partial x^{\mu}}{\partial x'^{\nu}}\,u'^{\nu}
	=\left(\frac{u'^{0}}{\sqrt{2}},0,0,\frac{u'^{0}}{\sqrt{2}}\right)\,.
\end{equation}
Using the relation $g_{\mu\nu}=e_{a\mu}e^{a}{}_{\nu}$ together with
Eq.~\eqref{static}, we obtain the following set of tetrads:
\begin{equation}
	e_{a\mu}=
\begin{pmatrix}\label{tetrads}
\frac{1-A^{2}}{A} & 0 & 0 & A \\
0 & 1 & 0 & 0 \\
0 & 0 & 1 & 0 \\
A & 0 & 0 & -A
\end{pmatrix}\,,
\end{equation}
where $A=-1/\sqrt{2-H}$.

In defining the photon energy, we project the photon four-momentum onto the temporal component of the observer tetrad, identified with the observer's four-velocity. 

From Eqs.~\eqref{enerproj} and \eqref{tetrads}, the photon energy measured
by the observer reads
\begin{equation}\label{energy}
	p^{(0)}=\frac{1}{\sqrt{2-H}}
	\left(
	\frac{\dot{x}^{2}}{2}
	+\frac{\dot{y}^{2}}{2}
	-\frac{H}{2}
	+1
	\right)\,.
\end{equation}
The spatial components of the photon momentum in the observer frame are
given by
\begin{equation}
	p^{(1)}=\dot{x}\,,
	\qquad
	p^{(2)}=\dot{y}\,,
\end{equation}
and
\begin{equation}
	p^{(3)}=\frac{1}{\sqrt{2-H}}(\dot{v}-1)\,.
\end{equation}
As a consistency check, one verifies that $p^{a}p_{a}=0$, confirming that
the observer measures the photon as propagating at the speed of light.

\begin{figure}[t]
 \centering
 \begin{minipage}{0.48\linewidth}
 \centering
 \includegraphics[width=\linewidth]{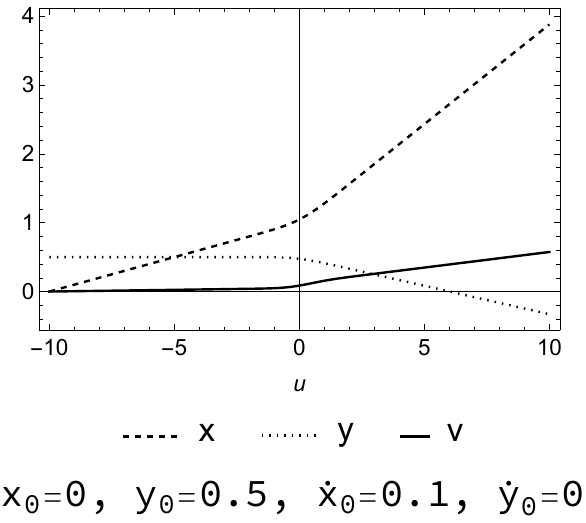}
 \par\smallskip
 {\small (a) Energy gain.}
 \end{minipage}\hfill
 \begin{minipage}{0.48\linewidth}
 \centering
 \includegraphics[width=\linewidth]{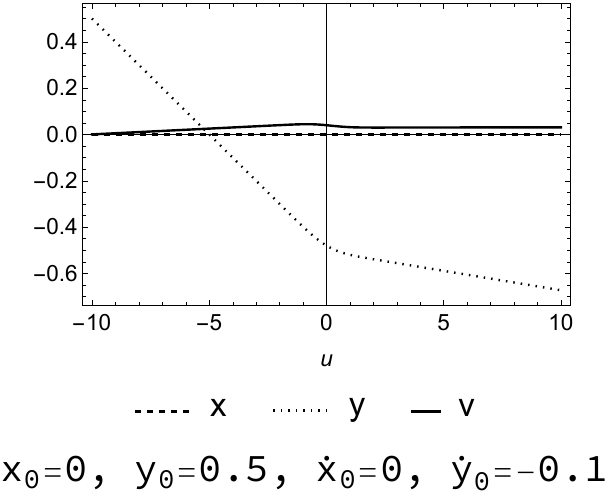}
 \par\smallskip
 {\small (b) Energy loss.}
 \end{minipage}
 \caption{Coordinate evolution for the $+$ polarization for the same initial transverse position.}
 \label{fig1}
 \end{figure}

\begin{figure}[t]
 \centering
 \begin{minipage}{0.48\linewidth}
 \centering
 \includegraphics[width=\linewidth]{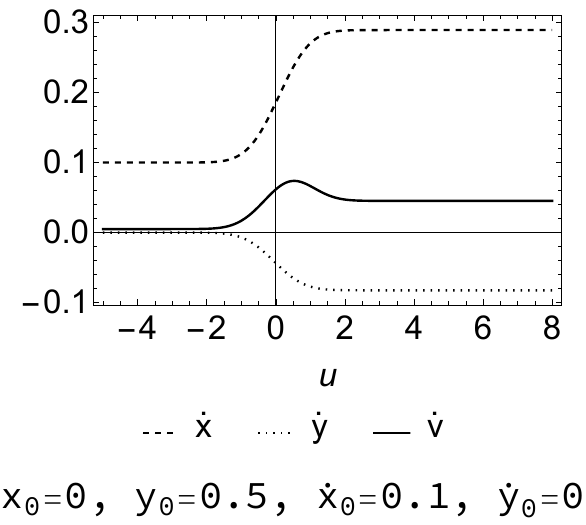}
 \par\smallskip
 {\small (a) Energy gain.}
 \end{minipage}\hfill
 \begin{minipage}{0.48\linewidth}
 \centering
 \includegraphics[width=\linewidth]{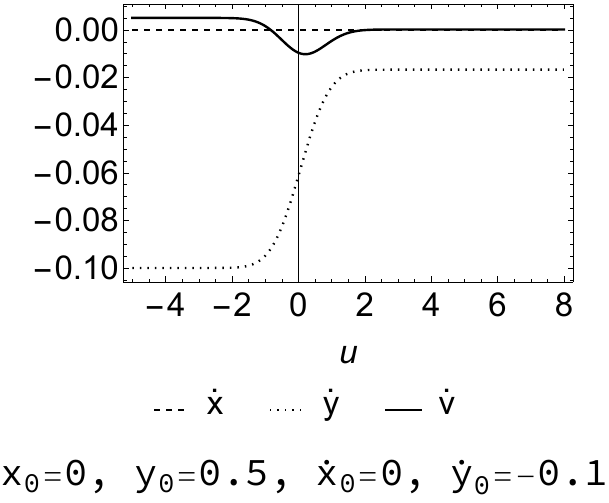}
 \par\smallskip
 {\small (b) Energy loss.}
 \end{minipage}
 \caption{Velocity evolution for the $+$ polarization for the same initial transverse position.}
 \label{fig2}
 \end{figure}

\begin{figure}[t]
 \centering
 \begin{minipage}{0.48\linewidth}
 \centering
 \includegraphics[width=\linewidth]{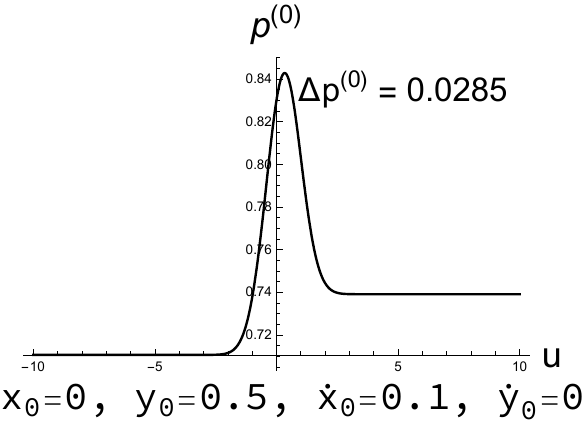}
 \par\smallskip
 {\small (a) Energy gain.}
 \end{minipage}\hfill
 \begin{minipage}{0.48\linewidth}
 \centering
 \includegraphics[width=\linewidth]{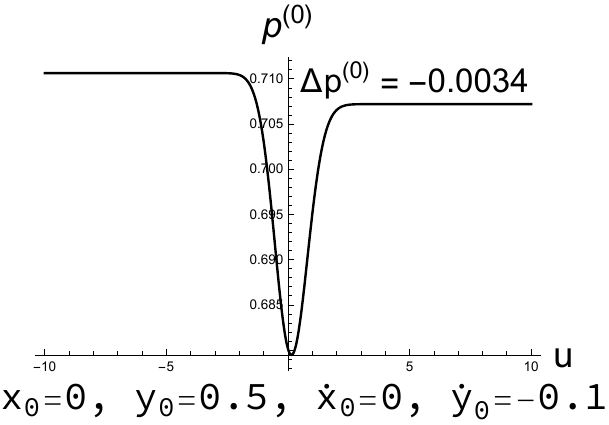}
 \par\smallskip
 {\small (b) Energy loss.}
 \end{minipage}
 \caption{Photon energy evolution for the $+$  polarization for the same initial transverse position.}
 \label{fig3}
 \end{figure}

The geodesic equations~\eqref{geotrans} and \eqref{geolong} are solved
numerically by specifying the polarization mode and the initial
conditions for the photons composing the light ray. We observe that the
qualitative behavior of the observer-measured photon energy depends
sensitively on the initial conditions. E.g., for the $+$
polarization mode, photons initially propagating in the positive $x$
direction exhibit an energy gain, whereas photons propagating in the
negative $y$ direction undergo an energy loss. This behavior is
illustrated in Figs.~\ref{fig1}--\ref{fig3}.

This behavior is not exclusive to the $+$ polarization. Similar qualitative
features are observed for other modes and polarizations, although not
necessarily for the same sets of initial conditions.

\subsection{Energy-gain and energy-loss regions}\label{sec3.1}

We aim to identify the conditions under which the photon energy associated
with a light ray increases or decreases. From Eq.~\eqref{cartesians}, in
the asymptotic limit $H=0$, one finds
\begin{equation}\label{energyassin}
	p^{(0)}=\frac{1}{\sqrt{2}}(\dot{v}+1)\,,
\end{equation}
and the total variation of the energy is given by
\begin{equation}\label{deltap}
	\Delta p^{(0)}=\Delta \dot{z}
	=\dot{z}_{f}-\dot{z}_{i}\,.
\end{equation}
From Eq.~\eqref{deltap}, it follows that, in order for the photons
composing the light ray to lose energy, their longitudinal velocity must
decrease (not necessarily in magnitude). Equation~\eqref{energyassin}
shows that the photon energy is directly related to the longitudinal
velocity component, so that an energy decrease is associated with regions
of the wavefront that reduce the longitudinal velocity of the photons.

Differentiating Eq.~\eqref{geolong} with respect to $u$, one obtains
\begin{equation}
\ddot{v}
= \frac{1}{2}\frac{dH}{du}
+ \dot{x}\,\ddot{x}
+ \dot{y}\,\ddot{y}\,.
\end{equation}
Substituting Eq.~\eqref{geotrans} into the expression above yields
\begin{equation}\label{vddot}
\ddot{v}
= \frac{1}{2}\,\partial_{u}H
+ \partial_{x}H\,\dot{x}
+ \partial_{y}H\,\dot{y}\,.
\end{equation}
For a separable profile of the form $H(u,x,y)=f(u)\,F(x,y)$, with $f(u)$
describing a localized pulse, Eq.~\eqref{vddot} becomes
\begin{equation}
\ddot{v}
= \frac{1}{2}f'(u)\,F
+ f(u)\,\frac{dF}{du}\,.
\end{equation}
The total variation of $\dot{v}$ is then given by
\begin{equation}\label{eqdeltav}
\Delta \dot{v}
= \frac{1}{2}\int_{-\infty}^{+\infty}
f(u)\,\frac{dF}{du}\,du\,,
\end{equation}
where an integration by parts was performed and the localization property
of the pulse, $f(\pm\infty)=0$, was used.

Equation~\eqref{eqdeltav} shows that when the light ray accesses regions in
which the quantity $f(u)\,\frac{dF}{du}$ is negative, the photons
composing the ray tend to lose energy, provided that the negative
contribution dominates the integral. Conversely, one may choose initial
conditions that favor either energy increase or energy decrease. Since
$f(u)>0$, a net energy increase occurs when the light ray explores regions
where $f(u)\,\frac{dF}{du}$ is positive, whereas a net energy decrease
occurs when the ray predominantly probes regions where
$f(u)\,\frac{dF}{du}$ is negative.

Using the relation $\Delta \dot{v}=\sqrt{2}\,\Delta \dot{z}$, one finally
obtains
\begin{equation}\label{main}
\Delta p^{(0)}=\Delta \dot{z}
= \frac{1}{2\sqrt{2}}\int_{-\infty}^{+\infty}
f(u)\,\frac{dF}{du}\,du\,.
\end{equation}

If the light ray initially propagates purely along the negative $z$
direction ($\dot{x}_{0}=0=\dot{y}_{0}$), its longitudinal velocity is
minimal and cannot decrease further. Consequently, according to
Eq.~\eqref{deltap}, the photon energy associated with the light ray can
only increase. In contrast, for an initially purely transverse
propagation satisfying $\dot{x}_{0}^{2}+\dot{y}_{0}^{2}=2$, a symmetry
emerges between the conditions for energy increase and energy decrease, as
inferred from Eq.~\eqref{main}. This suggests that light rays with a
larger longitudinal component tend to gain energy, whereas predominantly
transverse light rays tend to maintain energy balance after a finite
number of random interactions.

By combining these two limiting cases, one may conjecture that, after
several random interactions with multiple pp-waves, the average energy of
the photons composing a light ray increases. This conjecture is examined
quantitatively in the next subsection.

\subsection{Statistical tendency of energy variation}\label{sec3.2}

We investigate whether a light ray propagating over long distances
preserves its initial energy after successive interactions with multiple
gravitational waves. To this end, we consider ensembles of photons with
different initial conditions and analyze the statistical distribution of
their energy variation after the interaction with a pp-wave with polarization $+$. By comparing the number of
occurrences of energy increase and energy decrease, one may infer the
existence of a systematic tendency toward net energy gain or loss.

We first adopt a uniform sampling of initial conditions, with transverse
positions and velocities randomly drawn from the intervals
\begin{align}
	-1 \leq x_{0},y_{0} \leq 1\,,\nonumber\\
	-\sqrt{2} \leq \dot{x}_{0},\dot{y}_{0} \leq \sqrt{2}\,,
\end{align}
and with $v_{0}=0$. For a sample size of $n=5\times10^{4}$ photons, the
typical statistical uncertainty is
$\delta \simeq 1/\sqrt{n}\approx 0.4\%$.

The results for different wave amplitudes $A$ are summarized in
Table~\ref{tab:uniform}. As the amplitude decreases, the distribution of
energy variation becomes increasingly balanced, approaching
equipartition between energy gain and energy loss. Nevertheless, even for
moderate amplitudes, a statistically significant bias toward energy gain
is observed. This indicates that strong gravitational waves tend to
induce a net blueshift, while weak waves primarily preserve the photon
energy, apart from small fluctuations.

\begin{table}[t]
\centering
\caption{Statistical distribution of photon energy variation for uniformly
sampled initial conditions, with $n=5\times10^{4}$.}
\label{tab:uniform}
\begin{tabular}{c c c c}
\hline\hline
Amplitude $A$ 
& Energy gain (\%) 
& Energy loss (\%) 
& $\Delta p^{(0)}_{\mathrm{max}}$ \\ 
\hline
$10^{-1}$ & $86.53$ & $13.47$ & $5.71$ \\
$10^{-2}$ & $54.39$ & $45.61$ & $2.91\times10^{-1}$ \\
$10^{-3}$ & $50.57$ & $49.43$ & $2.68\times10^{-2}$ \\
$10^{-4}$ & $50.33$ & $49.67$ & $2.65\times10^{-3}$ \\
\hline\hline
\end{tabular}
\end{table}

We have also verified that this near-equipartition regime at small amplitudes is not restricted to the lowest polarization mode $m=2$: repeating the same analysis for higher-order profiles $H_{3R}$, $H_{3I}$, $H_{4R}$ and $H_{4I}$ yields comparable gain/loss fractions for $A=10^{-4}$, indicating that the statistical tendency reported in Table \ref{tab:uniform} extends beyond the lowest modes.

In addition to the uniform sampling, we consider a non-uniform ensemble
designed to mimic a more physical situation, namely photons with fixed
asymptotic energy and isotropically distributed propagation directions at
infinity. In this case, the initial transverse positions are uniformly
distributed over a disk in the transverse plane, while the transverse
velocities follow a rotationally invariant Gaussian distribution centered
at zero. The longitudinal velocity is determined by the null condition,
implying a kinematical lower bound that prevents photons close to
counter-propagation from further decreasing their energy.
This lower bound is not a fundamental property of the spacetime itself, but
results from the null condition together with the choice of observer frame
used to measure photon energy. Photons initially propagating opposite to the
wave direction already attain the minimal longitudinal velocity allowed by
the null constraint, so that any perturbation of their trajectory necessarily
increases $\dot{z}$ and therefore the measured energy.

Numerically, this non-uniform ensemble exhibits an even stronger
statistical bias toward energy gain when interacting with sufficiently
strong gravitational waves. As the amplitude decreases, this bias is
gradually suppressed, and the distribution of energy variation approaches
a symmetric, diffusive regime. These results indicate that, while no
universal pointwise energy gain exists, strong gravitational waves induce
a probabilistic tendency toward photon blueshift, whereas weak waves lead
predominantly to energy dispersion with negligible net drift.

We also evaluated significantly smaller amplitudes, down to
$A\sim10^{-14}$, to probe the weak-field regime closer to typical
astrophysical strains. In this limit the distribution of photon energy
variation approaches a nearly symmetric gain--loss balance, with only small
residual deviations within the numerical accuracy of the simulations.

The variations in photon energy induced by interactions with pp-waves are
inherently stochastic, depending on the initial conditions and trajectories of
the photons. From an observational perspective, if these effects are not
accounted for, they would manifest as additional noise in redshift measurements,
since standard analyses only consider known sources of uncertainty. This
highlights a potential methodological implication: unidentified contributions to
redshift may be mistakenly interpreted as experimental noise, even though they
reflect real, albeit theoretically predicted, physical effects.

\section{Conclusions}

In this work, we investigated the interaction between light rays and pp-wave spacetimes, focusing on the memory effect experienced by null geodesics and its consequences for photon energy. Motivated by ongoing discussions concerning tensions in precision cosmology and the interpretation of redshift measurements, we examined whether gravitational radiation can induce cumulative and permanent frequency shifts during photon propagation.

By numerically solving the null geodesic equations in pp-wave backgrounds, we showed that light rays approaching the wave from generic directions generically retain a memory of the interaction. This memory manifests itself both in the deformation of initially parallel beams and in permanent changes in the photon kinematical state. Except for the special case in which the photon propagates parallel and in the same direction as the wave, a memory effect is always present.

Using a coordinate-invariant definition of photon energy based on tetrads adapted to a stationary observer, we demonstrated that the interaction with pp-waves can lead to either energy gain or energy loss, depending on the initial conditions and the wave profile. While no universal pointwise energy shift exists, our statistical analysis reveals a robust tendency toward energy gain for physically motivated ensembles, particularly for photons propagating opposite to the wave direction. This behavior originates from the combined effect of the kinematical lower bound imposed by the null condition and the nonlinear transverse dynamics induced by the gravitational wave.

These results suggest that photon propagation through gravitational radiation may introduce an additional, path-dependent contribution to the observed redshift. From an observational standpoint, the measured redshift may be decomposed as
\begin{equation}
z_{\mathrm{obs}} = z_{\mathrm{cosmo}} + z_{\mathrm{pec}} + \delta z_{\mathrm{GW}},
\end{equation}
where $\delta z_{\mathrm{GW}}$ encodes the cumulative effect of interactions with gravitational waves and is intrinsically stochastic. Although subdominant with respect to the cosmological redshift, this contribution may accumulate over long propagation distances and naturally generate dispersion in redshift measurements.

The amplitudes considered in this work should be regarded as illustrative parameters of the pp-wave model rather than as direct representations of astrophysical gravitational-wave strains. 
The mechanism identified here
depends on the local interaction between the photon trajectory and the
gravitational wave. In practice, a sufficiently strong pp-wave could in
principle induce deviations in redshift measurements that might appear as
additional scatter if not properly accounted for, whereas weaker waves
would likely produce effects that remain observationally negligible.
Thus, the present analysis should be viewed primarily as a theoretical
framework for understanding how gravitational-wave memory effects could,
under suitable conditions, influence precision redshift measurements.

We emphasize that memory effects in plane and pp-wave spacetimes have been
extensively investigated in different contexts, including optical
properties, geodesic deviations, and kinematical memory for massive and
massless particles
\cite{Harte2015, DattaGuha2024, WangFeng2024}.
The present work differs from these analyses in that the memory effect is
characterized directly in terms of the photon energy measured by a
stationary observer, providing a physically operational definition of
gravitationally induced redshift and blueshift.
This formulation allows us to identify the conditions under which the
interaction of null geodesics with a pp-wave spacetime leads to a net
energy gain or loss as an asymptotic memory effect, establishing a direct
connection with redshift-based observational inference.

Although this effect shares the generic property of a permanent change of state
with the gravitational-wave memory described in \cite{Christodoulou1991} and \cite{StromingerZhiboedov2016}, our result differs fundamentally in terms of the
spacetime context, amplitude, and potential observability. The energy memory
analyzed here pertains to photons propagating through pp-waves, which are not
asymptotically radiative, and can produce variations in observed redshift that
might be interpreted as noise if unrecognized. This provides a theoretical
explanation for potential discrepancies in redshift measurements without assuming
the waves behave like linearized astrophysical gravitational waves.

The energy memory effect described in this work occurs in a theoretical
pp-wave spacetime but it provides a conceptual framework that could inspire tests in
suitable analog gravity systems, where frequency shifts and memory phenomena can
be studied under controlled laboratory conditions \cite{rozenman2024observation,tung2013gravitational}.

Therefore, gravitational memory effects provide a quantitatively well-defined relativistic mechanism that may contribute to small discrepancies and intrinsic scatter in astrophysical redshift data, without challenging the standard interpretation of cosmic expansion. Our results indicate that such effects should be regarded as a systematic correction in high-precision cosmological analyses, rather than as an alternative explanation for cosmic acceleration.


\bibliographystyle{unsrt}
\bibliography{bibitex}

\end{document}